# Physical Color Calibration of Digital Pathology Scanners for Robust Artificial Intelligence Assisted Cancer Diagnosis

Xiaoyi Ji[1], Richard Salmon[2], Nita Mulliqi[1], Umair Khan[3], Yinxi Wang[1], Anders Blilie[4,5], Henrik Olsson[1], Bodil Ginnerup Pedersen[6], Karina Dalsgaard Sørensen[7,8], Benedicte Parm Ulhøi[9], Svein R. Kjosavik[4,10], Emilius A.M. Janssen[4,5], Mattias Rantalainen[1], Lars Egevad[11], Pekka Ruusuvuori[3,12], Martin Eklund[1], Kimmo Kartasalo[1]

1. Department of Medical Epidemiology and Biostatistics, Karolinska Institutet, Stockholm, Sweden
2. FFEI Life Science, FFEI Ltd, Hemel Hempstead, UK
3. Institute of Biomedicine, University of Turku, Turku, Finland
4. Department of Pathology, Stavanger University Hospital, Stavanger, Norway
5. Faculty of Science and Technology, University of Stavanger, Stavanger, Norway
6. Department of Radiology, Aarhus University Hospital, Aarhus, Denmark
7. Department of Molecular Medicine, Aarhus University Hospital, Aarhus, Denmark
8. Department of Clinical Medicine, Aarhus University, Aarhus, Denmark
9. Department of Pathology, Aarhus University Hospital, Aarhus, Denmark
10. Faculty of Health Sciences, University of Stavanger, Stavanger, Norway
11. Department of Oncology and Pathology, Karolinska Institutet, Stockholm, Sweden
12. Faculty of Medicine and Health Technology, Tampere University, Tampere, Finland

Corresponding author: Kimmo Kartasalo, kimmo.kartasalo@ki.se.

The potential of artificial intelligence (AI) in digital pathology is limited by technical inconsistencies in the production of whole slide images (WSIs), leading to degraded AI performance and posing a challenge for widespread clinical application as fine-tuning algorithms for each new site is impractical. Changes in the imaging workflow can also lead to compromised diagnoses and patient safety risks. We evaluated whether physical color calibration of scanners can standardize WSI appearance and enable robust AI performance. We employed a color calibration slide in four different laboratories and evaluated its impact on the performance of an AI system for prostate cancer diagnosis on 1,161 WSIs. Color standardization resulted in consistently improved AI model calibration and significant improvements in Gleason grading performance. The study demonstrates that physical color calibration provides a potential solution to the variation introduced by different scanners, making AI-based cancer diagnostics more reliable and applicable in clinical settings.

# Introduction

Application of artificial intelligence (AI) to digital pathology has achieved highly accurate performance on tasks such as cancer detection and grading[1,2], prognostication[3,4], and prediction of molecular biomarkers[5]. However, it has been noticed that technical inconsistency during the scanning process introduced by hardware variance in light sources, sensors and quantizers[6–8], can degrade AI model performance[9,10]. Variations in image appearance can also be caused by different post-processing operations applied by the scanning software to the raw image data. This is a considerable hurdle to widespread clinical application of AI models, since tweaking of the models at each new site where they are to be deployed is not feasible. Site-specific fine-tuning, which is commonplace for currently commercially available algorithms, is also problematic in terms of medical device regulations and validation of the algorithms' real-world performance.

Even if an algorithm would have been fine-tuned to a given laboratory, unexpected behavior can arise due to changes in the imaging workflow, such as damage or wear of scanner components, or system updates of scanning software. In clinical use, this can lead to incorrect diagnoses and poses a threat to patient safety. Statistical methods, such as conformal prediction[11], have been proposed for detecting such changes in the data generation process, indicating that the outputs of the AI algorithm may no longer be reliable. While detecting such issues is important, avoiding them altogether with the use of robust AI algorithms and normalization or standardization techniques would be preferred. In the remainder of the paper, normalization is used to refer to approaches that reduce variation across whole slide images (WSI) in relation to an arbitrary reference among a given set of input data, while standardization produces image characteristics conforming to a universal reference standard.

The simplest, brute-force approach is to use training data from an increased number of sources to obtain improved generalizability of AI systems. However, it will be difficult to account for all the possible sources of variation that new scanner instruments may produce in the future. A widely applied solution for this problem is data augmentation, i.e. perturbation of the colors, contrast or other image characteristics randomly during training[12]. It is, however, challenging to design augmentations such that they are realistic and cover the real-world variance. Data augmentation operations devised to improve performance on a given dataset may not generalize to other data sources, where different types or magnitudes of variation may be present.

A computational stain normalization process for reducing the color variability between different scanners is an alternative way to address the problem[13]. A popular method proposed by Macenko et al.[14] estimates the haematoxylin and eosin (H&E) stain vectors of the WSI of interest by using a singular value decomposition (SVD) approach. Images with normalized color characteristics are then obtained by applying a correction to account for the stain intensity variations relative to a reference WSI[15]. This and other color normalization methods have produced mixed results[12]. For example, the application of color normalization by Duran-Lopez *et al.* did not lead to an advantage in prostate cancer diagnosis[16]. Another state-of-the-art color normalization method is the cycle-consistent generative adversarial network (cycle-GAN) algorithm, a technique that involves the automatic training of image-to-image translation models without paired examples[17]. The limitations of this method include the generation of unexpected artifacts and the high cost of frequent model retraining, which is required not only for every new data source but likely also periodically at a given site to account for variation in sample preparation and scanning over time. A further complication with most computational stain normalization methods is the need to select a representative reference WSI, or a set of WSIs, which is challenging to do in a well-defined, objective manner.

Computational normalization and augmentation of color also impart site-to-site bias on data, perturbing human validation of AI models by manipulating digital data away from the reality of glass tissue slides. As well as requiring large datasets to achieve statistical confidence and often further tuning to refine the normalization, making methods unreliable for most small research groups or local clinics, these methods do not provide simple, quantitative assessment against which non-computational users, like doctors or technicians, can assess the quality assurance (QA) of the diagnostic output and have confidence in the standard of results across their peers and patients.

It has been shown that utilizing a physical color calibration slide can standardize the color variations in WSIs arising from the scanning process[18]. According to the FDA, this slide should contain a set of measurable and representative color patches, and the color patches should have similar spectral characteristics to stained tissue[19]. With the help of this color calibration slide, one can calculate the International Color Consortium (ICC) profile of scanners[20] (https://www.color.org/specification/ICC.1-2022-05.pdf), and the color characteristics of each WSI can be calibrated to conform to a specific target ICC profile. In a pilot study, the application of a color calibration slide was shown to improve the visual concordance of WSIs to a microscope in over 50% of trials and improve diagnostic confidence by pathologists[6]. While calibration of WSIs to a particular known physical standard as such is not currently crucial for AI-based analysis, application of color calibration also provides a standardizing effect, removing some of the problematic variation in WSI characteristics and therefore reducing the burden on AI to solve variation with a traceable and explainable process.

We hypothesized that physical calibration of scanners can provide a consistent and lightweight means of standardizing WSIs to be used as input in AI-based analysis, resulting in improved generalization across different scanners and sites. While the computational pathology community has focused considerable efforts on studying and improving the ability of AI models to generalize to different settings, physical color calibration is, to the best of our knowledge, unexplored as a potential solution to the generalization problem. To test this hypothesis, we used as an example the task of prostate cancer detection and Gleason grading, which has attained considerable attention in recent years[1,9,21–23]. Here, we apply a commercial color calibration slide for standardizing WSIs of prostate biopsies collected at four different sites each using a different scanner model, and assess the effect of color calibration on the diagnostic performance of an AI system for detecting and grading prostate cancer. In addition, to benchmark the performance of the calibration slide against computational normalization methods, we applied the Macenko and cycle-GAN algorithms to the datasets and compared the resulting improvements to AI robustness.

## Methods

### Sample collection

To explore color calibration among WSIs created by different scanners, samples from four different sites were included in the study. For model training, we used prostate core needle biopsies from the STHLM3 clinical trial (ISRCTN84445406)[24] of 2012-2014, which was a prospective, population-based, screening-by-invitation study for evaluating a blood test based diagnostic model for prostate cancer in men aged 50–69 years residing in Stockholm, Sweden[1]. All the biopsy cores were graded based on the International Society of Urological Pathology (ISUP) grading classification by an experienced urological pathologist (LE), who also delineated the cancerous areas with a marker pen and measured the linear cancer extent. In total, 3651 biopsies from 957 participants were digitized with 20× magnification using an Aperio AT2 scanner (Leica Biosystems, Wetzlar, Germany) at SciLifeLab, Uppsala, Sweden.

For model tuning and evaluation, we obtained 329 WSIs from 73 men from Karolinska University Hospital, Stockholm, Sweden, scanned by a Hamamatsu NanoZoomer S360 C13220-01 scanner (Hamamatsu Photonics, Hamamatsu, Japan) with 20× magnification. Among them, 100 (30.4%) slides were assigned for tuning the hyperparameters of the prediction models and for training the GAN, and the rest for the evaluation of model performance. The samples were prepared in a different laboratory than the training set and thus represent an external test set. These slides were all graded by LE, providing a consistent diagnostication standard for the data used to train and test the model.

We also included WSIs from two additional external sites. Aarhus University Hospital, Aarhus, Denmark, offered 102 biopsy cores (42 men) imaged by a Hamamatsu NanoZoomer 2.0-HT scanner with 20× magnification, where 30 (29.4%) slides were assigned to a tuning set for training the GAN and 72 slides for a test set. From Stavanger University Hospital, Stavanger, Norway, 1228 WSIs from 200 men were obtained with a Hamamatsu NanoZoomer S60 with 40× magnification. Similarly, 368 (30.0%) slides were assigned into a tuning set to train the GAN and the rest were kept as a test dataset. These slides were diagnosed and graded by pathologists from the corresponding hospitals. Slides were assigned to the tuning and test sets randomly, stratified by ISUP grade. All the core needle biopsy slides in this study were routinely stained with haematoxylin and eosin (H&E) in the respective laboratory at each site.

The study protocol was approved by the Stockholm regional ethics committee (permits 2012/572-31/1, 2012/438-31/3, and 2018/845-32) and the Regional Committee for Medical and Health Research Ethics (REC) in Western Norway (permits REC/Vest 80924, REK 2017/71). Informed consent was provided by the participants in the Swedish dataset. For the other datasets, informed consent was waived by the institutional review board due to the usage of de-identified prostate specimens in a retrospective setting. The detailed composition of training, tuning and test data in terms of cancer length and cancer grade can be found in Table 1.

## Physical color calibration

For physical color calibration, we used the commercial calibration slide Sierra (FFEI Ltd., Hemel Hempstead, UK)[25]. The slide captures the color spectrum of commonly used histopathology stains identified from real and varied clinical samples. Briefly, when the slide was designed, mouse embryo tissue slides were stained using standard staining protocols as well as with each of the stains independently, and 300–400 spectral measurements from each slide were taken across each entire sample to generate color and intensity results representative of histopathology ground-truth. A biopolymer which performed well at preserving pathology stains with the same spectral response as stained tissue was identified for creating the color calibration slide. The main part of the slide consists of patches of biopolymer (Fig. 1) including H&E with varying combinations of stain intensities. The rest of the patches are stained with other common stains that cover the whole gamut of histopathology. To calibrate the color of a WSI, an ICC profile of the corresponding scanner is established using the color calibration slide. This involves measuring the known, true colors (CIELAB) on the color calibration slide using a calibrated spectrophotometer and then scanning it with a WSI scanner to generate RGB values containing scanner-induced variance from truth. Using a proprietary algorithm, intermediate color values for the ICC profile were accurately interpolated from the 'known' values from the color calibration slide.

For the purposes of this study, we scanned Sierra on each of the four scanners using imaging settings identical to the biopsy specimens. The same Sierra color calibration slide was scanned at Aarhus University Hospital in June 2020, SciLifeLab in September 2020, and Karolinska University Hospital in October 2020. At Stavanger University Hospital, scanning took place in November 2021, and a new Sierra slide was used to avoid effects due to aging of the calibration slide. An ICC profile was generated for each scanner at FFEI based on the four WSIs. As the target color profile, we used the standard sRGB ICC profile downloaded from https://www.color.org/srgbprofiles.xalter. We applied the ICC profiles to map images to a chosen standard color space using the *ImageCms* module (1.0.0 pil) from the *Pillow* library (8.0.0) in Python.

## Macenko color normalization

The color normalization approach was adopted from the work by Macenko et al.[14], with adjustments made to estimate slide level instead of patch level stain vectors. As the first step of the analysis, luminosity was adjusted per slide to enhance the stability of stain estimation. The process started with estimation of a luminosity reference, with a set of 100 random patches from the WSI under study sampled, concatenated and transformed from RGB to CIELAB color space. The slide level luminosity reference was then computed with the $95^{th}$ percentile ($I_{ref95}$) pixel values corresponding to the $L*$ channel. Next, the following steps were applied to correct the luminosity for each patch: firstly, the patches were converted to CIELAB color space; secondly, the pixel values lower than the reference were scaled linearly between 0 and 255, otherwise, set to 255; thirdly, tiles were transformed back to RGB space.

To estimate the reference stain vector ($V_{ref}$), 2000 luminosity corrected patches were randomly sampled across all WSIs in the training set. The patches were then appended and transferred from RGB ($I_{RGB}$) into optical density ($OD$) space [26] (equation (1)) to make the stains and saturations linearly separable. Pixels with an $OD$ value less than 0.15 were regarded as transparent and discarded from the analysis. Next, a plane spanned by the two vectors corresponding to the two largest singular values was formed with SVD on the concatenated patches. All pixels were subsequently projected onto the plane, followed by normalizing to unit length. Afterwards, the angles of each projected pixel to the first vector were calculated. Finally, the vectors corresponding to the $99^{th}$ percentile or $1^{st}$ percentiles among all angles were regarded as the H and E stain vectors, respectively.

$$OD = (-1 * log_{10}(I_{RGB}/255)) \qquad (1)$$

For each slide, a slide level stain vector ($V_{slide}$) was computed in the same manner as described above but with 100 randomly sampled patches within each individual slide.

With the estimated stain matrices ($V_{slide}$), the pixel level concentration coefficients ($S_{pixel}$) were computed with the $OD$ transformed pixel values ($OD_{pixel}$)[27] using the following formula:

$$S_{pixel} = V_{slide}^{-1} * OD_{pixel} \qquad (2)$$

The concentration estimates were also normalized by excluding one percent of the highest concentration values and then linearly rescaling the rest to match the pseudo maximum concentration value.

Lastly, color normalization was achieved by multiplying the reference stain ($V_{ref}$) with the normalized pixel level concentrations ($S_{pixel}$), and the output was eventually transformed back to RGB color space.

The code was adapted from the following packages: 'StainTools' (https://github.com/Peter554/StainTools), and 'Staining Unmixing and Normalization in Python' library (https://github.com/schaugf/HEnorm_python).

## GAN color normalization

Stain normalization of WSIs is a type of image-to-image translation. Generative adversarial network (GAN)[28] based methods have been adopted for image-to-image translation in recent years[17,29]. These methods are primarily divided into supervised and unsupervised categories according to the learning paradigm applied. Since paired data, required by supervised methods, does not exist in this type of stain normalization problem, we decided to use the unsupervised approach. There are several unsupervised GAN-based image-to-image translation methods[30–33] among which

CycleGAN[17] is the most commonly used. CycleGAN has previously successfully been used for histopathology image translation, including tasks such as stain-to-stain transformation[34,35], virtual staining of unstained tissue images[36–38], and stain normalization[39,40].

Our implementation of CycleGAN uses U-net-based[41] generators with skip connections and a PatchGAN-based discriminator[29]. The loss functions, optimizer, and their respective parameters are the same as in the original implementation[17]. We applied 50% augmentation of the data on the fly during training for each epoch by randomly flipping, rotating or scaling the patches. We used a batch size of 32 and the training processes were parallelized over 4 NVIDIA Volta V100 GPUs (Nvidia, Santa Clara, CA, USA) on the high-performance computing cluster Puhti (CSC - IT Center for Science, Finland).

Five data sets from four different sites, i.e. SciLifeLab (training), Karolinska (tuning, test), Aarhus (test), and Stavanger (test) were normalized using CycleGAN. First, a representative WSI was selected from the SciLifeLab data set (training) by a domain expert (LE), the WSI was then used as the target sample and the rest of the WSIs as source samples in the CycleGAN training setup. A separate GAN model was trained for each of the four sites based on a site-specific tuning set (see "Sample collection"). Models were trained from 10 to 25 epochs depending on the size of the data set. The epoch to use for each model was selected based on the generator loss and visual appearance of the validation results. To prevent any potential artifacts generated by cycle-GAN in the blank background of the patches, which had been observed in some cases, tissue masks were reapplied to all patches after being normalized by cycle-GAN, in order to mitigate the adverse effects of artifacts on the model performance.

## AI system

We based the AI system on a previously published model[1], where each WSI is processed by dividing the tissue sections into small patches. Firstly, a segmentation algorithm based on Laplacian filtering was applied to detect tissue (tissue mask) and annotations (pen mask). By projecting the pen mask onto the adjacent tissue, we created a label mask indicating cancerous and benign pixels. Patches of 598 × 598 pixels at a pixel size of 0.904 μm (around 540 × 540 μm) were extracted with 75% overlap and a minimum tissue content of 50%. Around 2.9 million patches were obtained during this process for model training.

For patch-level classification, we used two ensembles of 10 Inception V3 deep neural networks (DNNs)[42] each, pretrained on ImageNet. The first ensemble classified each patch into benign or malignant, and the second one graded a patch into Gleason patterns 3-5. Only patches from slides with one Gleason pattern (e.g. 3+3, but not 4+3)) were picked for training the grading ensemble to reduce label noise. The DNNs were trained by 10-fold cross-validation (CV) on the training dataset, resulting in one model per CV fold, and patch-level predictions were obtained for the out-of-fold WSIs of each model. Class balancing was applied by subsampling the majority classes on each epoch. For each patch, the ensembles output the probabilities of a patch being malignant and of a given Gleason pattern.

For aggregating patch-level predictions into WSI-level predictions of: 1) cancer presence, 2) cancer length, and 3) ISUP grade, we used gradient boosted trees implemented in XGBoost[43]. The three XGBoost models were trained using summary statistics of the patch-level predictions for each WSI as input, obtained from the DNNs via 10-fold CV on the training set. The patch-level predictions from the cancer detection DNN ensemble were used for the slide-level tasks of detecting cancer presence and estimating cancer length, and the predictions from the grading DNN ensemble were used for the task of slide-level ISUP grading.

To obtain predictions on test data, each WSI was first processed by the DNN ensembles to produce 10 sets of patch-level predictions, one per DNN model. Each set of patch-level predictions was summarized into slide-level statistics similarly as during training, and passed as input to the boosted tree models, producing one slide-level prediction of cancer presence, cancer length, and ISUP grade. The final predictions for each of the three outcomes were calculated by averaging the 10 predictions.

To optimize the numbers of epochs for training the two patch-level DNN ensembles, we saved the models every 10 epochs, training for a maximum of 100 (cancer detection DNN) or 30 (grading DNN) epochs, and chose the epoch that produced the best AUC for cancer detection and the epoch that produced the best Cohen's kappa for ISUP grading on the Karolinska University Hospital tuning dataset (see Statistical analysis for details). Two hyperparameters were tuned for the boosted tree models: the maximum depth of the boosted trees (max_depth) and the number of trees (num_rounds). For each of the three slide-level prediction tasks, we generated boosted trees with 40 combinations of these two parameters, where max_depth = 1, 2, ..., 8 and num_rounds = 100, 200, …, 500, and chose the models respectively with the highest AUC for cancer detection, the highest Cohen's kappa for ISUP grading and the highest linear correlation for cancer length estimation on the Karolinska University Hospital tuning data (see Statistical analysis for details).

All the training and evaluation jobs were run on high-performance GPU compute nodes on the clusters Alvis, part of the National Academic Infrastructure for Supercomputing in Sweden (NAISS), and Berzelius, at the National Supercomputer Centre (NSC).

## Statistical analysis

Cancer detection performance was assessed on slide-level with the area under the receiver operating characteristics (ROC) curve (AUC) and calibration curves. The calibration curves indicate the true frequency of the positive label against its predicted probability, for binned predictions, and were smoothed with a logistic regression.

To select an operating point for the cancer detection model, we picked the highest threshold which, when applied to the predicted probabilities, would result in no more than one false negative slide in the Karolinska University Hospital tuning set (sensitivity = 0.985). The same threshold was used later to evaluate slide-level sensitivity and specificity on all test datasets.

To assess cancer length prediction performance, we calculated the linear correlation between the cancer length estimated by the AI system and the cancer length described in the pathology report for each site and plotted the linear best fit line.

For slides predicted as positive for cancer, we further assigned the ISUP grade by applying the argmax rule on the predicted class-wise probabilities. We evaluated the performance of the AI system on ISUP grading using Cohen's kappa with linear weights against the pathologists' slide-level grading on the three independent test datasets. Linear weights emphasize a higher level of disagreement of ratings further away from each other on the ordinal ISUP scale, in accordance with previous publications[44].

All confidence intervals (CIs) were two-sided with 95% confidence level and calculated from 1000 bootstrap samples. DNNs were implemented in Python (version 3.6.9) using TensorFlow (version 2.3.0)[45], and all boosted trees using the Python interface for XGBoost (version 1.2.1).

## Results

We first assessed the ability of the AI system to discriminate between benign and malignant slides, first without any color management, and then with each of the calibration or normalization techniques applied to both the training and test data. All test data were external, representing samples prepared in a different laboratory and scanned on a different scanner than the training data. In the external test set from Karolinska Hospital, we observed AUC values of 0.989 (95% CI 0.973-0.999) and 0.971 (95% CI 0.942-0.992) with and without Sierra color calibration, respectively (Fig. 2A). Corresponding calibration curves are shown in Fig. 2D and Fig. 2G, respectively, indicating how well the probability of cancer presence estimated by the model corresponds to the true frequency of malignant slides across the range of predicted probabilities. Upon selection of an operating point for the cancer detection model, the resulting performance with and without Sierra is summarized in Table 2. The correlation coefficients between cancer lengths, as estimated by the AI system and the pathologist, exhibited a modest enhancement from 0.82 (95% CI 0.78-0.86) to 0.83 (95% CI 0.79-0.88) when Sierra calibration was applied (Table 3, Fig. S1). Regarding the performance of cancer grading, Cohen's linearly weighted kappa improved from 0.354 (95% CI 0.288-0.417) to 0.738 (95% CI 0.692-0.784) using Sierra (Table 4). To set this into context, the grading of the same samples by two different pathologists typically results in a kappa of 0.5-0.7[46].

We performed a similar analysis on the two other external datasets. On the Aarhus dataset, the AUC was 0.963 (95% CI 0.913-1.000) vs. 0.964 (95% CI 0.898-1.000) (Fig. 2B), the linear correlation was 0.82 (95% CI 0.68-0.91) vs. 0.81 (95% CI 0.71-0.89) and the Cohen's kappa coefficient was 0.452 (95% CI 0.346-0.557) vs. 0.423 (95% CI 0.319-0.541) with and without calibration, respectively. In the case of the Stavanger dataset, the machine learning model achieved an AUC of 0.982 (95% CI 0.972-0.990) vs. 0.976 (95% CI 0.963-0.986) (Fig. 2C), the linear correlation coefficients were 0.84 (95% CI 0.79-0.89) vs. 0.82 (95% CI 0.75-0.86) and the Cohen's kappa coefficient was 0.619 (95% CI 0.578-0.657) vs. 0.439 (95% CI 0.383-0.491) with

and without calibration. The performance is summarized for cancer detection, cancer length estimation and grading in Table 2, Table 3 and Table 4, respectively.

In addition, to compare Sierra with computational color normalization, we conducted otherwise identical experiments using Macenko and cycle-GAN methods for processing the training, tuning and test data. The results for the three external test sets are presented in Fig. 2, Fig. S1 and Table 2-4. Compared to the baseline of no normalization, the Macenko method provided a clear improvement on the Karolinska University Hospital data with AUC for cancer detection 0.983 (95% CI 0.966-0.996), linear correlation for cancer length 0.84 (95% CI 0.80-0.88), and kappa for Gleason grading 0.650 (95% CI 0.588-0.710). However, it led to slight deterioration of performance on Aarhus data with AUC 0.893 (95% CI 0.789-0.979), linear correlation 0.75 (95% CI 0.56-0.90), and kappa 0.408 (95% CI 0.295-0.513), and considerably degraded performance on Stavanger data with AUC 0.787 (95% CI 0.751-0.822), linear correlation 0.58 (95% CI 0.51-0.65), and kappa 0.281 (95% CI 0.221-0.340).

Cycle-GAN considerably improved performance on the Karolinska University Hospital dataset with AUC 0.992 (95% CI 0.985-0.998), linear correlation 0.89 (95% CI 0.86-0.92), and kappa 0.655 (95% CI 0.597-0.711). Its performance on the Aarhus dataset was mediocre, with AUC 0.918 (95% CI 0.840-0.981), linear correlation 0.72 (95% CI 0.51-0.88), and kappa 0.368 (95% CI 0.249-0.479). For the Stavanger slides, cycle-GAN performed comparably to the Sierra color calibration, with AUC 0.977 (95% CI 0.967-0.986), linear correlation 0.86 (95% CI 0.81-0.91), and kappa 0.623 (95% CI 0.578-0.663).

## Discussion

Our experiments demonstrated that the Sierra slide can serve as a robust physical color calibration method for AI-assisted computational pathology. In terms of the AI model's performance in discriminating between benign and malignant samples and assessing cancer length in the biopsies, physical calibration exhibited a consistent positive or neutral effect across different data sources. The advantage of applying Sierra became more apparent when examining the calibration curves (Fig. 2D-O). While the effect of physical color calibration on discriminatory capacity was rather modest, the improvement in AI model calibration was striking. This is reflected in the sensitivity and specificity of cancer detection measured at a classifier operating point that was specified on the tuning set, and then applied to the other cohorts without further adjustments (Table 2). For example, in the Stavanger University Hospital cohort, without any calibration the sensitivity of the AI model dropped to 51.8%. When Sierra calibration was applied, the sensitivity was retained at 90.4%, much closer to the intended tuning set value. Ultimately, samples typically need to be assigned a discrete classification decision (e.g. a cancer diagnosis or grade) as opposed to probabilistic outputs, and while often neglected in AI studies, model calibration is of crucial importance in view of practical applications[47].

Improved AI model calibration is further reflected in the considerably improved grading performance measured with the Cohen's kappa statistic (Table 4), which necessitates the choice of an operating point for the classifier and hence also measures model calibration. For example, the more than 100% improvement in Cohen's kappa on the Karolinska University Hospital cohort represents a difference between a diagnostic AI model that could be considered a risk to patient safety, and one that would perform comparably to pathologists. Crucially, unlike the computational color normalization approaches, applying physical color calibration consistently led to improved Gleason grading performance on all three cohorts. Even though physical calibration was not the top-performing method in all datasets when compared to computational color normalization approaches, the stable performance of the method is a crucial advantage and contrasts to the unpredictable behavior of computational normalization algorithms.

Computational stain normalization techniques have the advantage of being able to correct for variations arising from other sources than the scanner, but require considerable training or reference datasets. This is exemplified by the slight advantage of cycle-GAN normalization compared to Sierra calibration on the largest cohort of the study, Stavanger University Hospital. In contrast, for the Aarhus dataset, relying on only 30 slides to train the cycle-GAN, relatively weak performance was observed, particularly in terms of cancer grading. Moreover, while such models have the added capacity to adjust for staining variations, they risk generating image artifacts commonly referred to as 'hallucinations'[40,48], where the tissue morphology in the image is unintentionally altered. Similarly, for the more classical Macenko method, a representative sample of data from the target site is needed, which can pose issues scaling up to additional sites. In opposition to these methods, Sierra directly and specifically calibrates each image used, which may explain the consistent AI performance observed also for the smaller cohorts in this study. Indeed, this may provide a route to accurately 'scale-down' AI to sites producing less data such as research groups and local clinics, which in reality represent more numerous use cases than 'big data' from regional hospitals.

Besides the issues with scaling to an increasing number of sites, computational normalization methods pose challenges in the context of medical device regulations with respect to modification and validation of data and transparency of the processes being used. To account for changes over time in the sample preparation and scanning, approaches like the cycle-GAN would require retraining and revalidation of the model periodically at each site, which would be highly challenging in practice. However, Sierra provides independent quantification against a universal standard for real-time QA or GLP, using an established methodology that can be integrated anywhere in the post-scanning, pre-analysis workflow without altering the raw data, including integration into the scanners themselves. Taken together, our results suggest that physical color calibration can be a reliable approach for ensuring safe, quality-assured and robust deployment of AI algorithms across different clinical sites, using a standardized and simple method that does not require extra data-handling skills or site-specific tuning.

There are a number of limitations in the current study, which should be taken into account and addressed in future studies to further refine and enhance the effectiveness of physical color calibration. First, Sierra color calibration addresses the digital color fidelity differences between various scanners and effectively calibrates them to a standard based on spectral ground truth, but it does not account for the variations due to tissue processing and staining chemistry that occur before a glass slide is imaged. Still, despite the fact that Sierra only corrects for variation due to the scanner, it matched or outperformed the computational normalization methods included in this study. However, calibration could likely be further improved by developing physical techniques similar to Sierra, applicable to correct for variation due to tissue staining. Second, this study included four scanner models from two vendors, and it is likely that the impact of Sierra on AI performance would be greater with increased variability introduced by inclusion of more scanner vendors and new scanner models released over time. Third, there were time delays between the scanning of the Sierra and the actual slides used for evaluation. The time gap between scanning of the prostate samples and the Sierra slide was over one year on both the Hamamatsu scanner at Karolinska University Hospital and the Aperio scanner at Scilifelab. The positive impact of calibration observed in this study despite the considerable time delay provides some evidence of the stability of both the Sierra calibration method and the included scanners. Still, improved standardization accuracy is to be expected with prospective data collection and scanner profiling, as opposed to the retrospective profiling of Sierra in this study. Further investigation may be required to account for potential color fidelity discrepancies of the digital histopathology scanners over time as they drift from factory settings, and this may have implications for the recommended frequency of scanning Sierra slides in practical applications.

Gleason score is a strong prognostic factor for the survival of patients with prostate cancer and is crucial for treatment decisions[49]. Meanwhile, its subjective and non-specific nature is both a challenge and an opportunity for AI-assisted pathology. Consequently, the problem of AI-based Gleason grading of prostate biopsies has attracted considerable attention in recent years[23], and provided a relatively well-defined and algorithmically mature example application for the current study. Importantly, neither the problems with generalization of AI algorithms across different clinical sites and digital pathology scanners nor the color calibration technology evaluated in this study are specific to prostate cancer grading, and the methodology presented addresses a fundamental and universal issue in all WSI. To the best of our knowledge, this is the first study to evaluate the efficacy of physical color calibration as a potential solution to the problem of cross-site generalization and calibration of AI algorithms in computational pathology, and we expect subsequent studies to apply the same approach to other tasks, tissue types, and disease states.

There are fundamental task-dependent differences in the quality requirements that are to be placed on AI algorithms. For example, even though the recent breakthroughs with large language models have faced some skepticism due to the occasional factual errors committed by chatbots like Chat-GPT, individual erroneous responses typically do not bear similarly catastrophic consequences as in the case of an AI algorithm that clinicians rely on for medical diagnostics. If AI is to be widely applied in digital pathology, all aspects of the sample and data processing chain need to be scrutinized and quality-controlled rigorously, similarly to the processes applied to other medical measurement instruments[50] or even simple laboratory tools like pipettes. Ensuring consistent image quality is important also when the actual diagnostic task is performed by pathologists[51,52], but even more so if parts of the decision-making process are delegated to AI algorithms, which currently have limited capabilities to detect and report issues in their input data or to adapt to changes dynamically in a manner a human expert does. Errors conducted by machines are also often scrutinized more carefully and tolerated to a lesser extent than errors conducted by human experts, which sets the bar high in view of applying AI in medical diagnostics[53].

As a step in this direction, incorporating the use of a physical color calibration slide, such as the one presented in this article, could present a novel approach for efficiently enhancing the robustness, reliability and explainability of AI-assisted cancer diagnosis through histopathology images. A further positive impact is that the physical color slide methodology provides independent calibration and quantitative assessment against a known standard, allowing for inter-lab digital QA that will complement other QA tools for chemical staining variability as they become available. This would also be beneficial for data sharing, AI generalization as well as industry validation and regulation, especially if this method gains wider adoption. We believe that such improved techniques and processes for quality assurance should not be seen merely as incremental technical tweaks to existing AI methodology, but as the next essential step in translating AI algorithms from emerging technology into ubiquitous, routine clinical tools.

# Acknowledgements

R.S. received funding from Innovate UK (Future Leaders Fellowship MR/V023314/1). U.K. received funding from University of Turku (graduate school), Finland. A.B. received a grant from the Health Faculty at the University of Stavanger, Norway. B.G.P and K.D.S received funding from Innovation Fund Denmark (Grant no. 8114-00014B) for the Danish branch of the NordCaP project. M.R. received funding from Swedish Research Council and Swedish Cancer Society. P.R. received funding from the Research Council of Finland (Grant no. 341967) and Cancer Foundation Finland. M.E. received funding from Swedish Research Council, Swedish Cancer Society, Swedish Prostate Cancer Society, Nordic Cancer Union, Karolinska Institutet, and Region Stockholm. K.K. received funding from David and Astrid Hägelen Foundation, KAUTE Foundation, Karolinska Institute Research Foundation, Orion Research Foundation and Oskar Huttunen Foundation.

The computations were enabled by resources provided by the National Academic Infrastructure for Supercomputing in Sweden (NAISS) and the Swedish National Infrastructure for Computing (SNIC) at C3SE partially funded by the Swedish Research Council through grant agreements no. 2022-06725 and no. 2018-05973, and by the supercomputing resource Berzelius provided by National Supercomputer Centre at Linköping University and the Knut and Alice Wallenberg foundation (projects SNIC 2021/7-158, SNIC 2022/5-253 and Berzelius-2022-240).

We want to thank Carin Cavalli-Björkman, Astrid Björklund and Britt-Marie Hune for assistance with scanning and database support. We would also like to thank Simone Weiss for assistance with scanning in Aarhus and Silja Kavlie Fykse, Desmond Mfua Abono for scanning in Stavanger. ICC profile guidance was contributed by Craig Revie, John Stevenson-Hoare, Louise Collins and Jacqui Deane of FFEI Ltd. We would like to acknowledge the men that participated in the STHLM3 diagnostic study and NordCaP project, and contributed with the clinical information that made this study possible.

## Author contributions

Study concept and design: X.J., R.S., N.M., P.R., M.E., K.K.
Acquisition of data: A.B., B.G.P., K.D.S., B.P.U., S.R.K., E.A.M.J., L.E., M.E.
Analysis and interpretation of data: X.J., R.S., N.M., U.K., Y.W., M.R., P.R., K.K.
Drafting the paper: X.J., R.S., U.K., Y.W., M.E., K.K.
Critical revision of the paper for important intellectual content: X.J., R.S., N.M., U.K., Y.W., A.B., H.O., B.G.P., K.D.S., B.P.U., S.R.K., E.A.M.J., M.R., L.E., P.R., M.E.and K.K.
Obtaining funding: R.S., B.G.P., K.D.S., S.R.K., E.A.M.J., M.R., L.E., P.R., M.E., K.K.

## Competing interests

R.S. is an employee of FFEI Ltd. N.M. is an employee of Clinsight AB. Y.W. is an employee of Stratipath AB. M.R. is a co-founder and shareholder of Stratipath AB. L.E., M.E. and K.K. are co-founders and shareholders of Clinsight AB.

# Figures and Tables

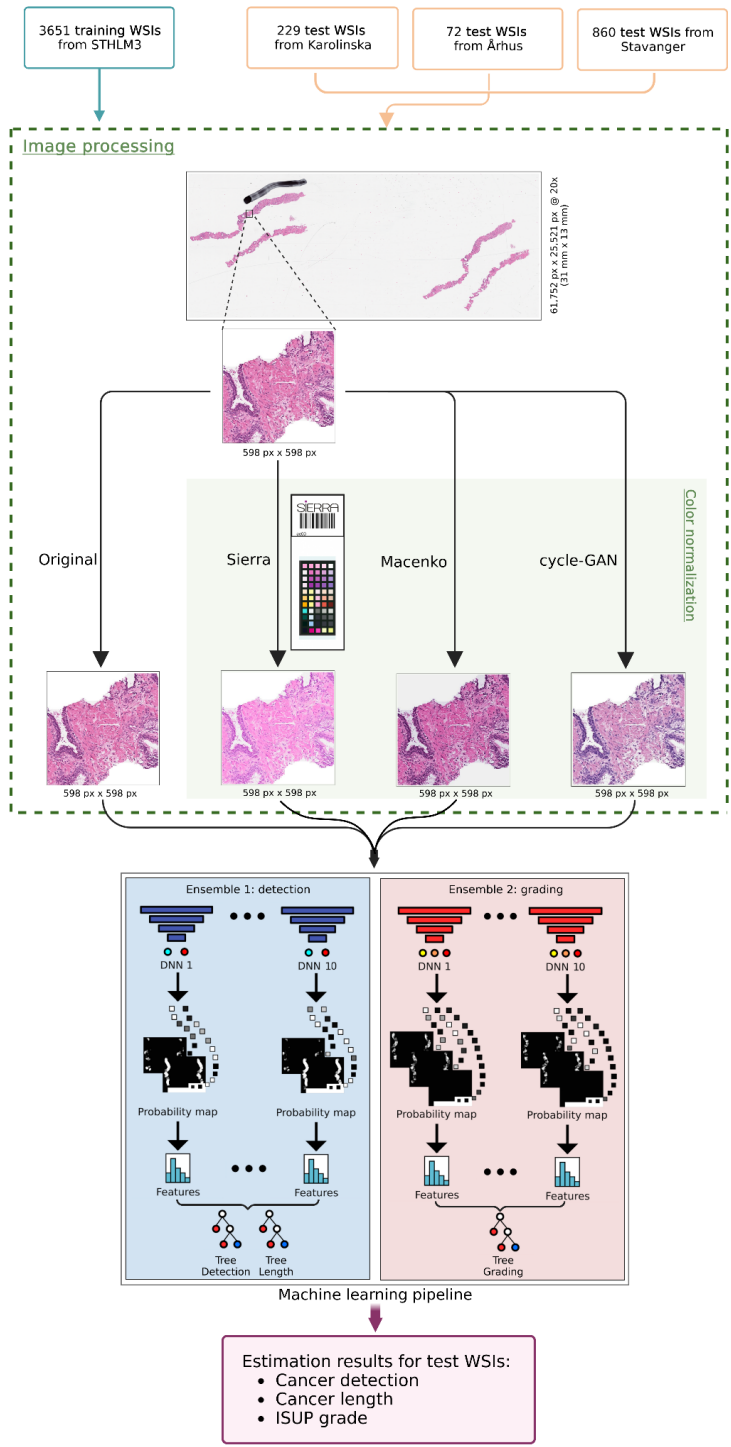

**Figure 1: Overview of the study pipeline.** The tissue region in the input WSI is first split into patches. Here, one patch is taken as an example for demonstrating the result of color calibration and normalization. The patches are fed as input to two model ensembles for cancer detection and cancer grading, which both consist of 10 deep neural networks for patch-level training and prediction, and gradient boosted trees for slide-level training and prediction. Models were trained independently on the original data, Sierra calibrated data, and data normalized with the Macenko or cycle-GAN algorithms.

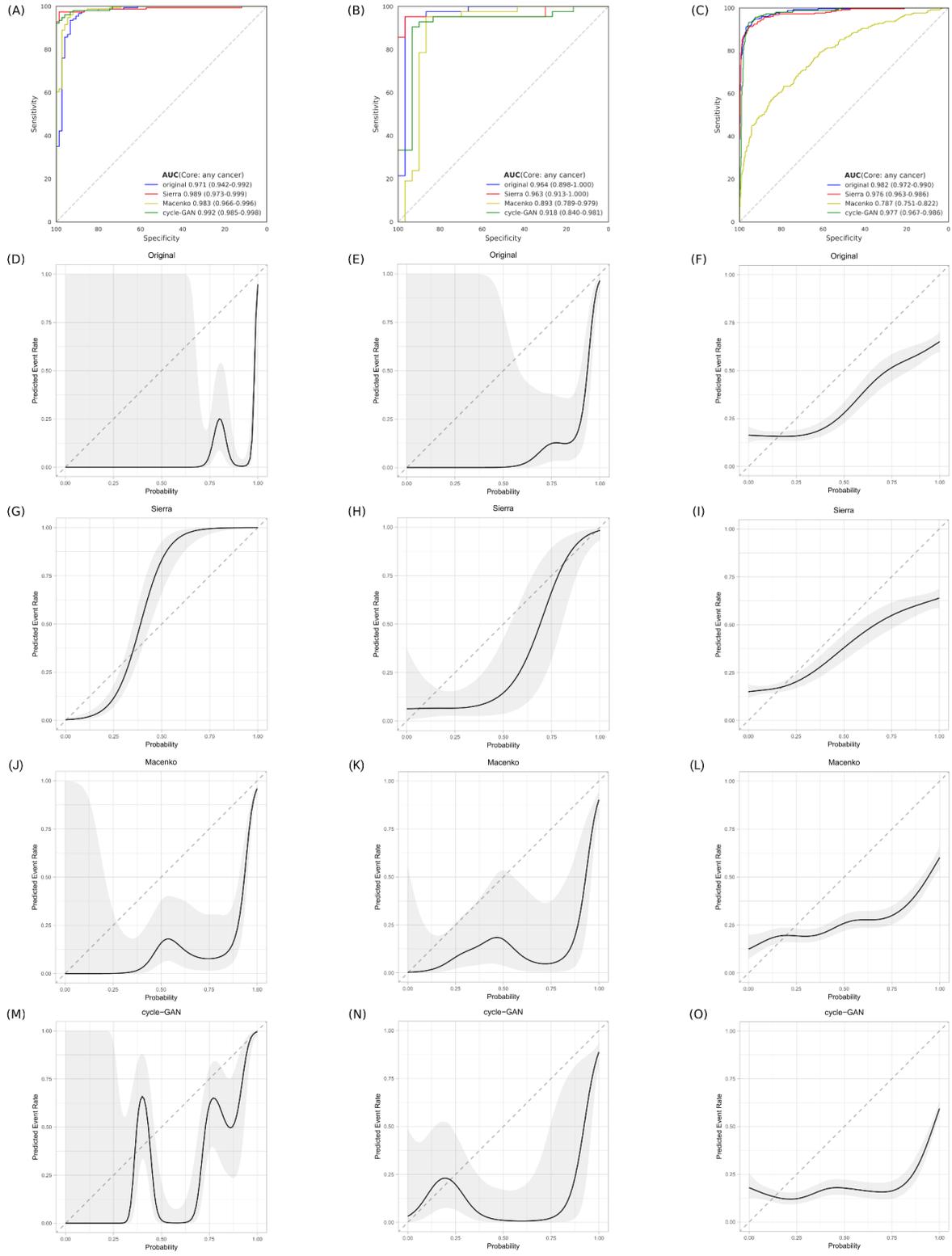

**Figure 2: Receiver operating characteristics curves with AUC (A~C) and calibration curves (D~O) for cancer detection in individual cores, with original, Sierra calibrated, Macenko and cycle-GAN normalized WSIs.** Columns from left to right: data from Karolinska University Hospital, Aarhus University Hospital and Stavanger University Hospital. Dashed gray lines in ROC curves represent the baseline curve corresponding to random guessing, while the ones in calibration curves represent an ideally calibrated model. The 90% confidence intervals for calibration curves are visualized using the gray ribbon. AUC = area under the curve.

**Table 1: Baseline characteristics of included prostate biopsy slides.**
The number of participants from each data source is indicated with *n*. ISUP = International Society of Urological Pathology.

| | Digitized biopsy slides | | | | | | |
|---|---|---|---|---|---|---|---|
| | STHLM3 (n = 957) | Karolinska University Hospital (n = 73) | | Aarhus University Hospital (n = 42) | | Stavanger University Hospital (n = 200) | |
| | Training | Tuning | Test | Tuning | Test | Tuning | Test |
| **Number of slides** | 3651 | 100 | 229 | 30 | 72 | 368 | 860 |
| **Scanner** | Aperio AT2 | Hamamatsu C13220 | | Hamamatsu C9600-12 | | Hamamatsu C13210 | |
| **Cancer length** | | | | | | | |
| No cancer | 739 (20.2%) | 33 (33.0%) | 75 (32.8%) | 13 (43.3%) | 30 (41.7%) | 261 (70.9%) | 609 (70.8%) |
| >0-1 mm | 752 (20.6%) | 8 (8.0%) | 24 (10.5%) | 3 (10.0%) | 2 (2.8%) | 24 (6.5%) | 55 (6.4%) |
| >1-5 mm | 1105 (30.3%) | 25 (25.0%) | 52 (22.7%) | 2 (6.7%) | 16 (22.2%) | 32 (8.7%) | 86 (10.0%) |
| >5-10 mm | 691 (18.9%) | 25 (25.0%) | 50 (21.8%) | 7 (23.3%) | 17 (23.6%) | 16 (4.3%) | 46 (5.4%) |
| > 10 mm | 364 (10.0%) | 9 (9.0%) | 28 (12.2%) | 5 (16.7%) | 7 (9.7%) | 35 (9.5%) | 64 (7.4%) |
| **Cancer grade** | | | | | | | |
| Benign | 739 (20.2%) | 33 (33.0%) | 75 (32.8%) | 13 (43.3%) | 30 (41.7%) | 261 (70.9%) | 609 (70.8%) |
| ISUP 1 (3+3) | 1156 (31.6%) | 19 (19.0%) | 45 (19.6%) | 8 (26.7%) | 18 (25.0%) | 62 (16.8%) | 145 (16.9%) |
| ISUP 2 (3+4) | 459 (12.6%) | 19 (19.0%) | 44 (19.2%) | 7 (23.3%) | 18 (25.0%) | 18 (4.9%) | 43 (5.0%) |
| ISUP 3 (4+3) | 309 (8.5%) | 15 (15.0%) | 34 (14.8%) | 0 | 1 (1.4%) | 13 (3.5%) | 32 (3.7%) |
| ISUP 4 (4+4, 3+5 and 5+3) | 576 (15.8%) | 6 (6.0%) | 18 (7.9%) | 2 (6.7%) | 5 (6.9%) | 7 (1.9%) | 15 (1.7%) |
| ISUP 5 (4+5, 5+4 and 5+5) | 412 (11.3%) | 8 (8.0%) | 13 (5.7%) | 0 | 0 | 7 (1.9%) | 16 (1.9%) |

**Table 2: Sensitivity, specificity and confusion matrix for slide-level cancer detection.**
Values in parentheses indicate 95% confidence intervals. The confusion matrices indicate true negatives (top-left), true positives (bottom-right), false positives (top-right) and false negatives (bottom-left).

|  | Original | | Sierra | | Macenko | | cycle-GAN | |
|---|---|---|---|---|---|---|---|---|
| **Karolinska University Hospital** | | | | | | | | |
| Sensitivity | 0.987 (0.966-1) | | 0.955 (0.917-0.987) | | 0.981 (0.955-1) | | 0.961 (0.929-0.987) | |
| Specificity | 0.813 (0.722-0.897) | | 0.987 (0.955-1) | | 0.88 (0.803-0.949) | | 0.96 (0.909-1) | |
| Confusion Matrix | 61 | 14 | 74 | 1 | 66 | 9 | 72 | 3 |
|  | 2 | 152 | 7 | 147 | 3 | 151 | 6 | 148 |
| **Aarhus University Hospital** | | | | | | | | |
| Sensitivity | 0.952 (0.875-1) | | 0.952 (0.875-1) | | 0.952 (0.881-1) | | 0.952 (0.880-1) | |
| Specificity | 0.967 (0.889-1) | | 0.933 (0.826-1) | | 0.867 (0.731-0.971) | | 0.833 (0.692-0.960) | |
| Confusion Matrix | 29 | 1 | 28 | 2 | 26 | 4 | 25 | 5 |
|  | 2 | 40 | 2 | 40 | 2 | 40 | 2 | 40 |
| **Stavanger University Hospital** | | | | | | | | |
| Sensitivity | 0.518 (0.456-0.580) | | 0.904 (0.867-0.938) | | 0.378 (0.316-0.443) | | 0.968 (0.944-0.988) | |
| Specificity | 0.997 (0.992-1) | | 0.959 (0.941-0.974) | | 0.954 (0.936-0.971) | | 0.882 (0.856-0.906) | |
| Confusion Matrix | 607 | 2 | 584 | 25 | 581 | 28 | 537 | 72 |
|  | 121 | 130 | 24 | 227 | 156 | 95 | 8 | 243 |

**Table 3: Linear correlations between cancer lengths estimated by the AI system and the pathologist.**
Values in parentheses indicate 95% CI.

|  | Original | Sierra | Macenko | cycle-GAN |
|---|---|---|---|---|
| **Karolinska** | 0.82 (0.78-0.86) | 0.83 (0.79-0.88) | 0.84 (0.80-0.88) | 0.89 (0.86-0.92) |
| **Aarhus** | 0.81 (0.71-0.89) | 0.82 (0.68-0.91) | 0.75 (0.56-0.90) | 0.72 (0.51-0.88) |
| **Stavanger** | 0.82 (0.76-0.86) | 0.84 (0.79-0.89) | 0.58 (0.51-0.65) | 0.86 (0.81-0.91) |

**Table 4: Linearly weighted Cohen's kappa for Gleason grading on test data.**
Values in parentheses indicate 95% CI.

|  | Original | Sierra | Macenko | cycle-GAN |
|---|---|---|---|---|
| **Karolinska** | 0.354 (0.288-0.417) | 0.738 (0.692-0.784) | 0.650 (0.588-0.710) | 0.655 (0.597-0.711) |
| **Aarhus** | 0.423 (0.319-0.541) | 0.452 (0.346-0.557) | 0.408 (0.295-0.513) | 0.368 (0.249-0.479) |
| **Stavanger** | 0.439 (0.383-0.491) | 0.619 (0.578-0.657) | 0.281 (0.221-0.340) | 0.623 (0.578-0.663) |

# Supplementary Figures

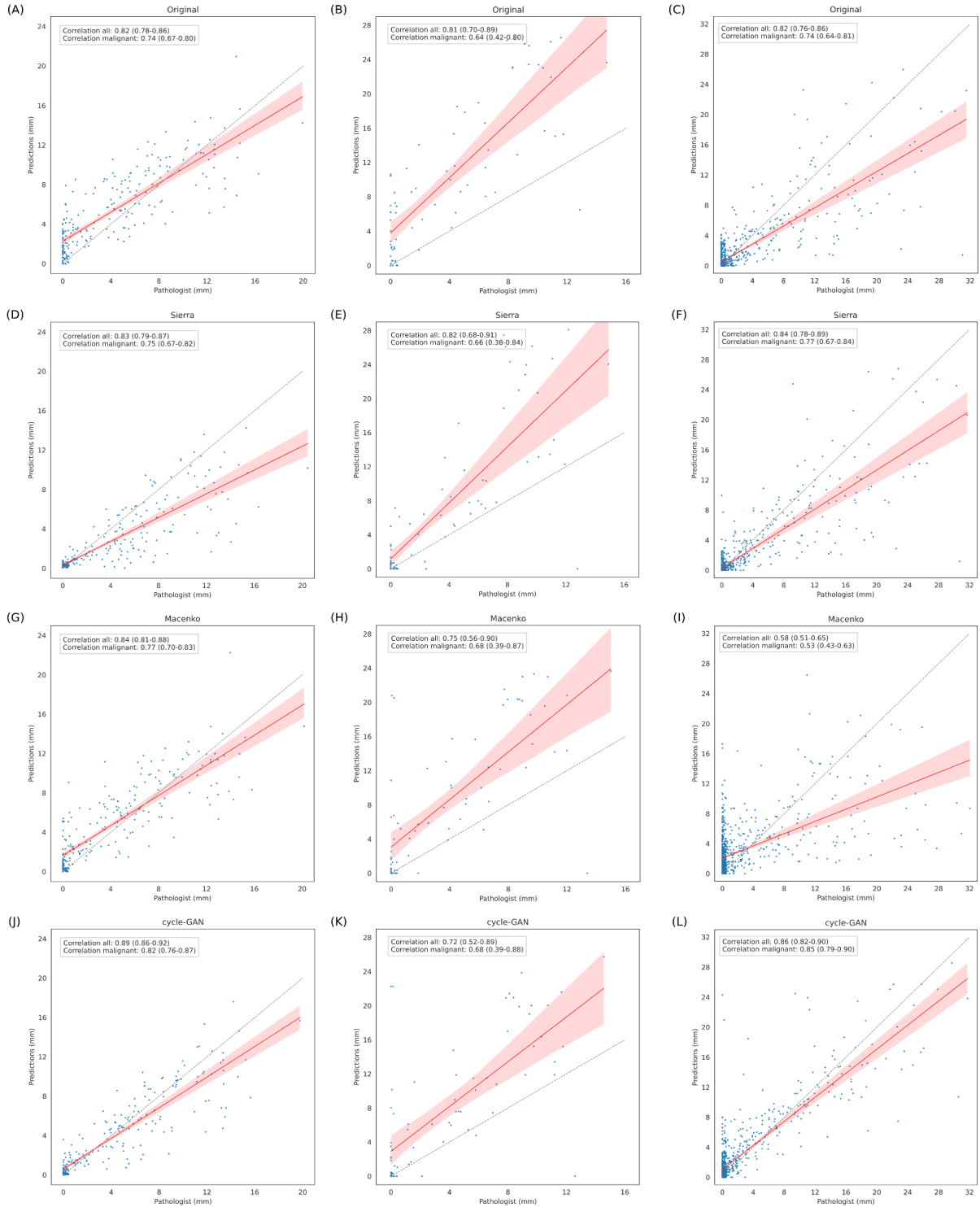

**Figure S1: Cancer lengths estimated by the AI and the pathologist, with original (A-C), Sierra calibrated (D-F), Macenko (G-I) and cycle-GAN (J-L) normalized WSIs.** Columns from left to right: data from Karolinska University Hospital, Aarhus University Hospital and Stavanger University Hospital. Linear correlation coefficients computed for all cores and malignant cores only are shown in each plot with 95% CI. Red lines and shading indicate linear best fits with 95% CI for each plot. Dashed gray lines indicate the ideal result.